\documentclass[pdftex]{sajlX}

\usepackage{latexsym, amssymb, amsfonts,amsmath}
\usepackage{graphicx}

\newcommand{\negr}[1]{\boldsymbol{#1}}

\usepackage{amssymb}

 \usepackage{adjustbox} 
\usepackage[english]{babel} 
\usepackage{verbatim}
\usepackage{latexsym}
\usepackage{graphics}
\usepackage{epsfig} 
\usepackage{graphicx}
\usepackage{fancybox} 
 \usepackage{ifsym} 
 \usepackage{multirow,bigdelim} 
 \usepackage[cmtip,arrow]{xy} 
 \usepackage{pb-diagram,pb-xy} 
\usepackage{comment} 
\usepackage{cjhebrew}

\def\rr{\noindent - }

\def\myal{ \hspace{-0.4ex}\raisebox{0.6ex}{\tiny \textit \textbf {)}}\hspace{-0.6ex} }   

\def\myaj{ \hspace{-0.4ex}\raisebox{0.6ex}{\tiny \textit \textbf {(}}\hspace{-0.6ex} } 

\newenvironment{myquote}%
  {\list{}{\leftmargin=0.15in\rightmargin=0.15in}\item[]}%
  {\endlist}

\title{Karl Pearson, Spinoza \& Causal Understanding} 
 {Karl Pearson and the Logic of Science:  \\  
  Renouncing Causal Understanding \\ 
  (the Bride) and Inverted Spinozism \\}

\author{Julio Michael Stern}{\mbox{} \\ 
	    Julio Michael Stern \\ 
        South American Journal of Logic, 
        v.4, n.1, pp.219-252, 2018. \\ 
        \mbox{} \\ }

\begin{document} 		  

\maketitle

\begin{abstract}
Karl Pearson is the leading figure of XX century statistics. 
 He and his co-workers crafted the core of the theory, methods 
and language of \textit{frequentist} or classical statistics 
 -- the prevalent inductive logic of contemporary science.    
  However, before working in statistics, K.Pearson had other 
interests in life, namely, in this order, philosophy, physics, 
and biological heredity. 
 Key concepts of his philosophical and epistemological  system 
of anti-Spinozism (a form of transcendental idealism) are 
carried over to his subsequent works on the logic of scientific discovery. 

This article's main goal is to analyze K.Pearson early philosophical and theological ideas and to investigate how the same ideas came to influence contemporary science, either directly or indirectly -- by the use of variant theories, methods and dialects of statistics, corresponding to variant statistical inference procedures and their specific belief calculi. 
     
\end{abstract}



\mbox{}  \vspace{2mm} \mbox{} 

\begin{flushright}

\textit{ 
 Scientific research can reduce superstition by encouraging people to \\ 
think and view things in terms of cause and effect. Certain it is that a \\ 
conviction, akin to religious feeling, of the rationality and intelligibility \\ 
of the world lies behind all scientific work of a higher order.  This firm \\ 
belief, a belief bound up with a deep feeling, in a superior mind that \\ 
reveals itself in  world of experience, represents my conception of God. \\ 
In common parlance this may be described as ``pantheistic'' (Spinoza).} \\ 
		 \mbox{} \vspace{-3mm} \mbox{}  \\  
		Albert Eistein, Ideas And Opinions, (1988, p.46, 262). \\

\pagebreak  

\<l:kAh dwodiy liq:ra't k*alAh>  \\ 
\<yA,siy,s `AlayiK: 'E:lohAyiK: \hspace{3.5mm}   
   k*im:,sU,s .hAtAN `al k*alAh> \\  
\<mero'+s miqEdEM n:sw*kAh \hspace{1mm} 
   swoP ma`a:,sEh b*:ma.ha:+sAbAh t*:.hilAh>  \\ 	
\mbox{} \vspace{-2mm} \mbox{} \\ 
{ Let's go, beloved, let's meet the bride:  \\  
Your God will rejoice over you, 
As the groom rejoices over his bride; \\ 
Before the beginning she was offered,   
Planned from start, Created at end.  \\ 
 \mbox{} \vspace{-3mm} \mbox{}  \\  
 XVI century verses by 
 Rabbi Shlomo Halevi Alkabetz \mbox{}  \\ }  
		     
\end{flushright}


\section{Introduction}  

Karl Pearson (1857-1936) is the leading figure of XX 
century statistics. Under his direct tutelage or influence, 
George Udny Yule (1871-1951),  Ronald A. Fisher (1890-1962), 
Jerzy Neyman (1894-1981),  Egon S. Pearson (1895-1980), 
and many others defined the methods, language and  
epistemology of the  frequentist school of mathematical statistics, 
that is, of XX century mainstream statistics -- the prevalent 
belief calculus used for inductive reasoning in the practice 
of  contemporary science.    
 This formalism translates K.Pearson's epistemological and 
philosophical position on the logic of scientific discovery, a position he calls:  
``Inverted-Spinozism --  a Spinozism modified by Fichte''.

Today, K.Pearson is known for his work in Statistics. 
However, his intellectual journey has many roots. 
Before working in Statistics, he studied and worked in 
philosophy, theology, physics, heredity and eugenics.  
 Moreover, his work in these fields had a compelling 
influence in his later work in statistics.    
Nowadays, many didactic text-books present K.Pearson's 
philosophical positions very unfaithfully, 
in watered-down pseudo-positivist or  
in sterilized decision-theoretic versions. 
Like the roots of a tall tree, the origins of  K.Pearson's philosophy 
are now deeply buried and often concealed, 
Nevertheless, like the branches of the same three, its epistemological and 
methodological consequences  are clearly seen and its influence 
strongly felt all over the realm of contemporary statistical science. 
 Moreover, since statistical test of hypotheses became the 
\textit{de facto} standard for validating scientific research, 
this influence is exerted much further.    
In this and following articles we analyze the historical and conceptual 
development of K.Pearson philosophy, and consider how it has 
influenced statistical inference procedures and the logic  
of corresponding belief-calculi. 
  
	The main goal of this paper is to present Pearson's epistemological 
	views, as he presented them in academic articles and books, and 
	also as presented allegorically in his novel \textit{The New Werther}.  
	Section 2 presents my interpretation of K.Pearson's  mystical journey 
	of  ``Renouncing the bride'', as narrated in \textit{The New Werther}. 
 Section 3 details more technical aspects of  his philosophical 
concept of Inverted-Spinozism. 	
 Sections 4 and 5 are conceived as a bridge, offering a connection  
to K.Pearson ideas concerning ether physics, heredity and eugenics, 
examining how these ideas were strongly influenced by his previous 
work in philosophy, and how they directly influenced his later work 
in statistics. 
 Following articles will discuss in detail the concepts of 
direct vs. inverse probabilities, analyzing its historical roots, 
K.Pearson's ideas on this topic, and their profound influence  
in XX century mathematical statistics. 
 Section 6 presents some final remarks, including other 
directions for future research, some conclusions  and 
acknowledgments.

\section{Sophia, Maimonides, Spinoza, Fichte, and Locki} 
  
		 K.Pearson arrives at his philosophical concept of Inverted-Spinozism 
	after a  religious  and spiritual crisis, and presents his positions in 
	several forms, including: His influential book 
	\textit{The Grammar of Science} (1892); 
	some review articles about the philosophy of Spinoza 
	(1880, 1883)  and, most importantly in our context; 
  his novel \textit{The New Werther -- by Locki} (1880). 
   This novel presents his philosophy in a romantic and mystical context, 
	that we explore in Section 2.2  
	via Spinoza's  philosophical roots in Jewish philosophical and 
	mystical writings by  Moshe ben Maimon (1135-1204), 
 Abraham Abulafia (1240-1291) and 
Joseph Gikatilla (1248-1310).  
 	Section 2.3 presents some of the motivations and some epistemological 
consequences of K.Pearson's Inverted-Spinozism.   	
		
\subsection{Love and Sex (or lack thereof) in \textit{The New Werther} }

K.Pearson describes the style of \textit{The New Werther} as	schw\"{a}rmerisch romanticism, using the term  Schw\"{a}rmerei in the sense of enthusiastic fervor. In German, this term usually means rapture or infatuation, and it is also used in religious literature  to denote zealotry, in the sense of utmost  commitment to ethical values 
and self-renunciation, see Levine (2010).  
 In the next subsections, I present a philosophical and 
mystical interpretation of this outstanding and unusual work, that I see 
as an act of true philosophical and spiritual renunciation, namely,  
\textit{Renouncing the Bride}, as explained in the sequel.  
I also explain why I see \textit{The New Werther} as the point of departure 
and the fulcrum  for all of K.Pearson latter works in science and statistics.   
Nevertheless, the reader should be aware that my interpretation differs 
from existing ones, specially from that of K.Pearson's main biographer, 
Theodore Porter (2004, Ch.3, p.43-68). 

\textit{The New Werther} describes the journey of Arthur, a young 
Englishmen in Germany,  his adventures visiting Heidelberg university,  
and his wandering about the scenic routes of the Black Forest. 
The novel's full title, \textit{The New Werther -- by Locki},  reveals its fictional 
author, Locki -- named after the shape shifter, trickster and subversive  
Germanic  god --  that plays the role of a Faustian Mephistopheles, the scatterer and forgerer of lies,  
\<mepi.s t*opEl +sAqEr>,  as in Job 13:4.   
This epistolary novel involves three main characters: 
Arthur;  Ethel, Arthur's fiance; and Raphael, Arthur's friend 
at the university. 

Arthur unequivocally represents K.Pearson himself.  
According to Porter (2004, p.54-55) well documented interpretation, 
Raphael is a fictional character partially based on Raphael Wertheimer, 
K.Pearson's friend and Law student at Heidelberg. 
Furthermore,  Porter (2004, p.55-56) concludes that  ``the role of Ethel 
was an elaboration of Pearson's friendship with Robert Parker'',  
K.Pearson's study partner and colleague at Cambridge.  
Furthermore Porter considers their relationship suggestive of the 
``upper-class male relationships at school in the Victorian era, 
and the efflorescence of homosexual activity at King's College'', 
pondering however that ``there is no evidence to support, and much 
reason to doubt, that this friendship was ever explicitly sexual''. 
  
In the next subsections I offer an alternative reading of 
\textit{The New Werther}, including an alternative interpretation of 
the characters Ethel and Raphael.  
I do not deny the validity of Porter's interpretation, not least because 
I do not have access to K.Pearson correspondence and other sources 
used by Porter in his biographical studies. 
 Moreover, I do not consider these interpretations to be mutually exclusive. 
Perhaps, all sexual tensions so accurately alluded by Porter could be 
considered as earthly affairs contrasting, and in this way complementing 
and highlighting,  the spiritual and philosophical  concerns discussed 
in the next subsections. 
After all, in the human soul ($\psi \upsilon \chi \eta$, Psyche), 
several aspects of sex, love and thought and  their representations appear 
to be deeply intertwined, as suggested by the ancient Greek myths of  
$\epsilon \rho \textrm{o}\varsigma$, Eros, in his two forms:  
either the playful young god of love, or his older form, 
$\phi \alpha \nu \eta \varsigma$ -- Phanes, the shiny god of revelation.  
Hence, it should  not be surprising to see all these aspects emerge 
together in the turmoil of K.Pearson's religious and spiritual crisis.

\subsection{Spinoza's Epistemological Principles and its Precursors}

\textit{The New Werther} presents K.Pearson's doctrine of 
Inverted-Spinozism in the aforementioned 
schw\"{a}rmerisch-romantic style, and is 
permeated by mystical elements. 
As an example, let us quote from the first encounter 
of Arthur and Raphael: 

\begin{myquote} 
\textit{ I was in the library, looking for a copy of 
Maimonides, to whom my study of Spinoza had 
led me, when I found a queer-looking person 
busily reading in a corner the very book I wanted. 
 He had a distinctly Jewish face, and yet beneath 
his raven-black and straggling locks there was an 
apostolic nobility and depth. He seemed to know 
by inspiration what I had come for, and with a 
salutation offered me the book.}
Pearson (NW, p.22). 
\end{myquote}

K.Pearson structures his philosophical doctrine of inverted-Sponizism around 
(and against) three major principles of Spinozian philosophy and epistemology, 
namely: \\  
(1)  \textit{Deus sive natura}; \\ 
(2) \textit{Cognitione causae} and 
      \textit{Leges naturae universales};  \\ 
(3)  \textit{Amor Dei intellectualis}. 

These epistemological principles are presented in axiomatic form at 
Baruch Spinoza's magnum opus, namely, his 
\textit{Ethica -- Ordine Geometrico Demonstrata} of 1677. 
K.Pearson had great familiarity, and also strong opinions 
about Spinoza's works and his commentators, even publishing in 
1880 a review of  \textit{Pollock's Spinoza} at the 
\textit{Cambridge Review}. 
K.Pearson was also well aware of precursors of these epistemological 
principles rooted in medieval Jewish philosophy,  as clearly attested 
by his article \textit{Maimonides and Spinoza} published at  
\textit{Mind} in 1883.    
 As examined in this section, these medieval philosophical precursors 
are expressed in a language and permeated by forms of argumentation 
that are perceived by contemporary readers as essentially mystical 
in nature,  for further comments see Stern (2017) and also 
Fraenkel (2006), Harvey (2007) and Idel (2000).  
  
In our context, the mystical route to Spinoza's epistemological principles 
offers some advantages over the axiomatic (ordine geometrico) exposition, 
among others: 
(a) It is in the best spirit of \textit{The New Werther}, 
a main object of study in this article;  
(b) It offers efficient short-cuts, allowing for close and effective 
connections with and between topics concerning causal analysis in 
natural science to be studied in the next sections. 
Having the aforementioned objectives and justifications in mind, 
let us explore Spinoza's epistemological principles via the 
Maimon-to-Spinoza mystic route. 

\subsubsection*{Deus sive natura} 

Spinoza's principle of \textit{Deus sive natura} -- 
God, or equivalently, nature -- finds a precursor  in 
Moshe ben Maimon (1135-1204) classical formula: 
\textit{ha-pe{\myaj}ulot ha-{\myal}elohiyoth}, 
 \textit{ha-pe{\myaj}ulot ha-teb{\myaj}ayoth},  
\< hap:`ulwot ha':Elohiywot hap:`ulwot ha.tEba`ywot>; 
meaning: In the world we live in --  
Actions of God (are just) actions of nature.  

Abraham Gikatilla (1248-1310) explains Maimonides' formula 
using forms  of  sylleptic argumentation that are typical of 
Jewish medieval philosophy and mysticism, see Stern (2017). 
His argument is based on the careful examination of 
the \<+sorE+s>, \textit{ shoresh}, root or stem \<.tb`>. 
On one hand, this stem generates the word    
\<.tEba`>,    \textit{ teba},  a substantive meaning nature or substance;  
On the other hand, this stem generates the word  
\<.taba`>, \textit{  tab{\myaj}a},  a verb meaning  
sank, stamped, coined or  formulated;  including the derived form    
\<ma.t:b*e`a>,  \textit{  matbe{\myaj}a}, coin, type, formula.    
 Also of interest is the expression  
  \<.t:biy`Ut--`ayin>, \textit{ tebiyoth--{\myaj}ayn},  
meaning  intuition or insight,	literally, eye-impression;  
see Klein (1987, p.239-240, 337. 
 Gikatilla conclusion from this philological analysis is that ``things'', 
as they present  themselves in nature, are coined, patterned, 
and also behave according to  ideal types or abstract  formulas, 
types and formulas that convey the will of God. 
	 
	Interestingly, the  Latin words 
	\textit{causare, causa},  (to) cause;  are  etymologically related to   
\textit{cudere,  cusum}, (to) strike, hammer, forge, stamp, coin; 
suggesting an analogy that parallels Maimonides formulation and its 
later interpretation by Gikatilla. 
The same analogy is repeated in the Portuguese (the language of 
Spinoza's country of origin)  words  \textit{causa/ cousa}, 
the cause (of something)/ (some)thing (caused).

\subsubsection*{Cognitione causae and Leges naturae universales}

The preceding analysis of Spinoza's first epistemological principle, 
Deus sive natura, brings us directly to his second principle -- 
{The understanding of causes and universal natural laws}. 
In the next paragraphs we reproduce this principle  
in abridged form, for the full version see  
Ethics (1677, Part I, Axiom 4; Pref. for Part III \& IV).

\begin{myquote} 
So, the cause or reason why God, or nature, acts, and the  
cause or reason why He exists, are therefore one and the  
same. ...  The eternal and infinite Being,  which we call God or   
Nature,  acts by the same necessity as that whereby it exists.  
 ...  The knowledge of an effect involves and depends on  the  
knowledge of its cause.  ... and therefore, one and the same  
should be the method of understanding  the nature of all things  
whatsoever, namely, through nature's universal laws and 
rules.\footnote{\textit{ 
Ratio igitur seu causa, cur Deus seu natura agit et cur existit,  
una eademque est. ...  Aeternum namque illud et infinitum ens,  
quod Deum seu  naturam appellamus, eadem, qua existit  
necessitate agit. ...   Effectus cognitio a cognitione causae  
dependet et eandem involvit. ...   atque adeo una eademque  
etiam debet esse ratio rerum qualiumcumque naturam  
intelligendi, nempe per leges et regulas naturae universales.} }     
\end{myquote}

\subsubsection*{Amor Dei intellectualis}

Once the first and second principles are stated  
and accepted, one question presents itself: 
Can mankind, even if in approximate form, have knowledge of 
the ideal types or abstract formulas that regulate nature? 
Spinoza's third principle is to state the answer in the affirmative, 
this statement being also an affirmation of God's love. 
  From ancient times to medieval Jewish philosophy and 
kabalistic exegesis, this affirmative principle of God's love 
has been associated to a rich imagery based on Genesis 28:12:

	\begin{center} 
	\<waya.ha:loM  w:hineh sulAM  mu.sAb  'ar:.sAh  
	     w:ro'+swo  mag*iy`a    
	   	ha+sAmAy:mAh	 w:hineh mal:'a:key  'E:lohiyM 
			 `oliYM  w:yor:diyM  b*:wo;>  
	\end{center} 		
		
	\begin{myquote}      
   \textit{He [Jacob] had a dream, a ladder was set on the earth 
	with its top reaching to heaven; 
	and behold, the angels of God 
	were ascending and descending on it.}  
	\end{myquote} 
	
Abraham Abulafia, in his book Light of the Intellect  (1285), 
\<'Or ha,sEkEl>, \textit{Or ha-Sekhel},  analyzes  Maimonides 
philosophy in general and this third epistemological principle in 
particular, and reiterates the nature of the ascending and descending 
movements at Jacob's ladder, namely,  this (e)motion is love! 
Love of Divine insight, 
\textit{ahabah elohut sikhlut}, 
\<'ahabah  'E:lohUt  ,sik:lUt>, 
meeting the love of human understanding,  
\textit{ahabah enoshut sikhlut}, 
\<'ahabah  'E:nwo+sUt ,sik:lUt>.

Furthermore, in his book Life of the Soul (1275), 
\<sepEr .hayey hanEpE+s>,  \textit{sepher chayei  hanephesh}, 
Abulafia compares the joy of this Divine and human 
intellectual encounter at Jacob's ladder to the 
delight, \textit{tha\myaj anug}, \<t*a`a:nUg>, 
of the groom and the bride, 
\textit{ ha-chathan ve-ha-kalah}, 
\<h.hat*aN whk*alAh>.

Abulafia's romantic analogy conceals sylleptic arguments typical 
of Jewish medieval philosophy and mysticism, see Stern (2017). 
On one hand, the verb \<.hOteN>, \textit{chothen}, (to) marry, 
also carries the meanings of (to) circumcise and (to) join, connect.  
On the other hand, the word for bride, \<k*alAh>,  
also means -- to be completed,  that is related to the roots     
\<kol>, all, whole; \<k*Alal>,  (to) complete, perfect, generalize;  
and \<k*:lAl> general rule, principle or law.\footnote{
Gesenius (1906, p.368, 477-483) and 
Klein (1987, p.237, 276-278) offer etymological analyses 
based on classical methods of comparative linguistics. 
Clark (1999, p.118-119 p.293-301), Gesenius (1910, p.99-102), 
Frajzyngier (1979) and  Horowitz (1993)  
suggest possible derivations based on formation pathways 
from biliteral bases to triliteral roots, or  from a geminate root 
to its gradational variants.}

More detailed discussion of the topics presented in this subsection 
can be found in Fraenkel (2006), Harvey (2007) and Idel (2000). 
These articles give very erudite views of these topics, commenting 
several aspects beyond the scope of the present paper. 
For example, some of these articles show how to rephrase 
philological and sylleptic arguments into gematrical 
calculations, a recourse typical of medieval kabalah.

\subsection{Renouncing the Bride: Werther and Mephisto}  
		  
After studying in depth the philosophy of Spinoza and its precursors, 
K.Pearson begins to depart from its core principles, finally arriving  
at his ``Inverted-Spinozism -- a Spinozism modified by Fichte''. 
As he so honestly acknowledges, these modifications take him so far 
away from Spinoza that their philosophies become, in many aspects, 
antithetic; hence the name Inverted-Spinozism. 
 More technical aspects of this philosophical position are analyzed in 
Section 3, see also Jacobi (1994, p.502), Limnatis (2008, p.119-120) and 
Schelling (1994, p.107-108). 
 In this section we pay close attention to \textit{The New Werther} 
narrative. 

 My reading of \textit{The New Werther} tells a tale of  a groom, 
Arthur -- representing K.Pearson,  that rejects his bride.   
In this reading, Raphael becomes a spokesman for the 
philosopher Baruch Spinoza, while Ethel, Arthur's beloved bride, 
becomes a representation of  Greek Sophia 
($\sigma \textrm{o} \phi \iota \alpha$, wisdom) or Hebrew 
Shekinah (from the root \<+skn> shochen, to dwell within --  
meaning Divine understanding as it can and does dwell within a 
human being),  that is often symbolically represented in the 
kabalistic literature as the `cosmic bride'.  
That is, Ethel becomes a representation of  human profound 
understanding or intuitive insight that can be achieved by God's grace.   
 In his scientific and philosophical papers, K.Pearson rejects 
the possibility of this kind of understanding, whereas,    
 in \textit{The New Werther}, Arthur rejects the love of Ethel, doing so in 
the most dramatic way and with the most tragic consequences.

The extent to which K.Pearson accepts ``Inverted Spinozism''   
and, therefore, breaks away from the love of Ethel-Sophia,     
with all its philosophical and spiritual consequences, 
becomes clearer as we read \textit{The New Werther}  
according to my proposed interpretation. 
 The following key passages, 
 that I for convenience have labeled (i) to (vii), 
provide stepping stones for reading \textit{The New Werther}  
according to this interpretation. 
  
 In the first quotation, (i), Arthur  presents his reasons 
for staying away from his (formerly) beloved Ethel. 
 His motives still have an exploratory or tentative character, 
but his arguments will mature as the novel progresses, 
strengthening the determination of his decisions, 
and also the gravity of its consequences.

\begin{myquote}  		
(i)  \textit{ 
Let us show that the two great schools of materialism 
and idealism, which have divided the world against 
itself, are really at one; that the inexorable laws 
under which science asserts that the universe must 
for ever roll on, are not empiric, but deducible from 
the pure reason; and that, though the sway of the 
intellect shall thus be extended from the logical to 
the empirical, yet that the intellectual, the manly, 
shall itself be so bound up with feeling, the womanly, 
that the two shall be united in one being and in 
one life, as we have been. Let us prove that the 
Ideal is not a world invented by the painter and 
the poet, but that it exists in every Actual; that 
the Deity is not a cause outside and separate from 
material man; that the cause must not be sought 
outside, but rather in the effect -- nay, perhaps, is 
that effect itself. \ \ 
With this end, best of friends {\em [Ethel, Sophia]},  
have we renounced each other.}  
(NW, p.11). 		
\end{myquote}

The next set of quotations, labeled (ii) to (v), give key 
snapshots of this novel climax, turning points and anti-climax.   
 In the next scene of interest, he Germans are celebrating 
Walpurgisnacht, also known as Hexennacht -- the  night of the witches. 
 At this ancient Germanic pagan festival, celebrated around  
bonfires at the hilltops, Arthur has a cathartic moment. 
   At (ii), Arthur (K.Pearson) tries to convince Raphael 
(Spinoza) to turn his philosophy upside-down, to replace the  
absolute God by an absolute Ego -- an Ego that is 
capable of ripping the veil from the face of nature. 
  Raphael's answers (iii) with sarcastic questions: 
How could a mere mortal take command of the universe?  
How could a finite human being acquire infinite knowledge? 
How could such a claim be something more than arrogant hubris?  	
	Arthur is incapable of answering the unanswerable, 
but promises to seek justification for his position within  
the theoretical framework of German Idealism. 
  As an indication of Pearson's commitment to German culture  
it is interesting to mention that, during his trip to Germany, he 
replaced his English given name by its German version -- 
from Carl to Karl. 
  Finally, at (v), Arthur expresses a feeling of guilt, 
as if he had done something terribly wrong:   
 His heart senses he betrayed the love of Ethel (Sophia),  
and anticipates the dangers that lay ahead;  
  Nevertheless, his mind is already committed to follow
the way of departure, renouncing his promised bride.

\begin{myquote} 		
(ii)  \textit{Do you not feel, said I  {\em [Arthur, to Raphael]},  
on a night like this, a titanic fire burning in your soul; that 
Nature is for the moment, your slave; that the Spirit of the 
Universe is at hand, and that you can compel it 
to raise its veil?}  
\vspace{1mm} 

(iii)  \textit{
Raphael:  But  how can you, who  believe in the  
pantheism of Spinoza, allow that the part can possibly compel   
the whole; that man,  an infinitesimal portion  of the Godhead,  
can command the Spirit of the Universe?}  (NW, p.25-26). 
\vspace{1mm} 

(iv) \textit{ I [Arthur] have determined to read those German   
reasoners,  and then again to discuss the matter with Raphael  
 [Spinoza].}  (NW, p.28).  
\vspace{1mm}

(v)  \textit{The morning was cool, yet my head felt on fire.   
 It seemed to me as if someone had told me I had wronged  you,  
 Ethel [Sophia]; and I knew it was not true.}  
 (NW, p.29).  
\end{myquote}   

A few pages later, Arthur presents the core of his philosophy 
in a concise and well articulated form. The similarities with 
Fichte's philosophical position and the influence of the 
Inverted-Spinozism metaphor, as defined by Jacobi and 
expanded by Schelling, are fully acknowledged. 
  
\begin{myquote} 
(vi) \textit{ 
 The spirit, to be an ``I'', must have a nature of its own; 
a nature denotes constancy; constancy we term a law. 
The I, in order to exist, must follow this law; 
the permanency of forms which this law gives 
to our perceptions we term space and time. 
The thought-law which dictates permanency to the 
perceptions, compels us to look upon space and 
time as infinite and eternal. Thus we see how 
Eternity and Infinity, which stood before as twin 
giants, laughing at and ready at once to crush and 
obliterate the feeble actions of a finite humanity, 
are but the necessary creations of our own inner 
nature.}    Pearson (NW, p.39).  
\end{myquote}

Nothing good will come from following the advice of 
Locki, the trickster, no light will be found at the end of 
a path set by the father of Hell.  
 Arthur is no longer interested in Rafael's friendship, 
and suggests that Raphael makes a visit to Ethel, 
who lives in Paris, the city of light (and love). 
 For the next months, Arthur wanders trough the black forest 
in an individuation journey of sorts,  that he makes having as 
sole companion his loyal dog, Gaspar, see 
Franz (1990, p.137; 2000, p.59).  
  
 Emerging from the forest, he goes to Paris, only to find 
Ethel (Sophia) and Raphael (Spinoza) deeply in love 
with each other.  
 Arthur has an enraged tantrum, and after an outburst 
of jealousy and a melodramatic speech of false forgiveness, 
commits suicide.  
	At the last pages of \textit{The New Werther}, its fictional author, 
Locki,   assumes the narrative, for poor Arthur is no longer in 
condition to fulfill this role. 
	The very last words of \textit{The New Werther} bring Locki's  
concluding message (vii), that he delivers pronouncing the 
words of  the character Mephistopheles in Goethe's Faust:

\begin{center} 
(vii) \textit{ Ich bin der Geist, der stets verneint!   \\ 
 Und das mit Recht;  denn alles, was  entsteht,   \\ 
 Ist werth, da{\ss} es zu Grunde geht; \\ 
 Drum besser w\"{a}r's, da{\ss} nichts entst\"{u}nde.} \\    		
	{ [I am the spirit that all negates! \\ 
And rightly so, for all that originates, \\ 
Should rightly to its destruction run; \\  
'Twere better then that nothing were begun.] }     		
\end{center}

\section{Inverted Spinozism: Science for the  
         Absolute Ego} 
		
 In order to fully understand the consequences of the alternative 
interpretation of \textit{The New Werther} presented in the last section, 
we have to examine the origin and interpretations of the expression  
``Inverted Spinozism'', an expression coined and used by 
Friedrich Heinrich Jacobi (1743-1819) to criticize,  and also describe 
-- in a nutshell,  the philosophy  of Johann Gottlieb Fichte (1762-1814). 
 The same expression is recycled by K.Pearson to label his own 
philosophical position  as ``Inverted-Spinozism 
 -- a Spinozism modified by Fichte''. 
 This expression appears for the first time in an open letter from 
Jacobi (1799) to Fichte  from which, in the sequel, we quote some 
key paragraphs in abridged form, see Jacobi (1994, p.501-502): 

\begin{myquote}  
 \textit{ 
``I am'' and ``There are things outside me''. 
Speculative philosophy had to try to subordinate one of these propositions
to the other; to derive the former from the latter or the latter from
the former -- exhaustively, in the end -- so that there would be but one
being and one truth before its eye, the all-seeing one!   
... 
Thus the two main avenues, materialism and idealism, or the attempt
to explain everything from a self-determining matter alone or
from a self-determining intelligence, have the same aim. Their opposing
courses do not take them apart at all, but rather bring them gradually
nearer to each other until they finally touch.}  

 \textit{ 
Strange, that the thought never occurred to Spinoza
of inverting his philosophical cube; of making the upper side, the side
of thought which he called the objective, into the lower, which he
called the subjective or formal; and then of investigating whether his
cube still remained the same thing; still for him the one and only true
philosophical shape of reality. 
Everything would have transformed itself without fail under his hands 
at the experiment. 
 The cube that had hitherto been ``substance'' for him 
 -- the one matter of two entirely different beings -- 
would have disappeared before his eyes, and in exchange a
pure flame would have flared up, burning all by itself, with no need 
of place or material to nourish it: Transcendental Idealism!    
 ... 
I chose this image because I first found entry into the Doctrine of
Science {\em [Wissenschaftslere (1813)]} through the representation 
of an inverted Spinozism. } 
 \end{myquote} 
 
The expression Inverted-Spinozism was further disseminated by 
Friedrich Wilhelm Joseph von Schelling (1775-1854), and by several 
commentators ever since; 
In Schelling (1827, 1994, p.108) we find the following extended 
explanation of why Fichte's philosophy can be well characterized 
as an Inverted Spinozism:

\begin{myquote} 
\textit{ 
There was no alternative there, if one did not wish to move once again 
into the absolute object, witch destroys everything free in the subject, 
than to move to the opposite -- to the all-destroying subject, which was 
now no longer the empirical subject of Descartes, but only the absolute 
subject, the transcendental I.} ...   
  
\textit{  
 Since this I was not the empirical I, then for Fichte the {\em I am}, 
which he made into the highest principle of philosophy, could also 
not be an empirical fact -- Fichte declared it to be an action 
(Thathandlung, literally `deed-action') and showed how the I could 
in no way exist independently of this action as a dead, immobile 
{\em thing}, but only in this act of self-positing [Selbstsetzung]  
in which he recognized not just a temporal, and also not just a 
transitory beginning which had begun the movement at some time, 
but the beginning which was always equally eternal -- 
thus that, wherever and whenever one wanted to begin, 
this act of self-positing always had to be the beginning.}    
\textit{  	
 Fichte's idealism thus is the complete opposite of Spinozism or is an 
{\em inverted} Spinozism, because it opposes to Spinoza's absolute 
object, which destroys everything subjective, the subject in its 
absoluteness, opposes the {\em deed [Thathandlung, deed-action] } 
to the merely immobile being (seyn) of Spinoza;  the I is for Fichte, 
not as it is for Descartes, just something assumed for the purposes 
of philosophising, but the real, the true being, the absolute 
{\em Prius} of everything.} 
\end{myquote}

 The next sub-sections examine how K.Pearson faithfully and fully 
accepts the implications of being a philosopher of Inverse-Spinozism.   
 Sections 4, 5, and following articles will examine how he carried these 
positions to his way of doing science in general and statistics in particular. 
  The next subsections focus on two specific aspects of great importance, 
namely, the epistemological or ontological status of  the   
external world, and the role (or lack thereof) of causal 
or metaphysical explanations  in the practice of science.

\subsection{Human Mind and Reality of  ``the''  External World:  \\
 A Contrast with the Constructivism of Paul Volkmann}

Pearson restates the same views expressed at the last quotations 
of   \textit{The New Werther} using a strictly academic voice, 
see next quotation. 
		
\begin{myquote} 		
\textit{ We require in fact a kind of  inverted Spinozism, 
a Spinozism modified   by Fichte.  
  ... 
 There is no need of the pre-established Harmony 
  ...  
 The outer world, as we conceive it, is the production 
of the conceiving `Ego', not an objective reality enforcing 
its laws upon the subjective sensitive centre. 
 When we talk about a law of physical nature, we only mean 
a necessary law of thought, any such law is only an 
intellectual law, the necessary method in which we are 
compelled to view our sensations. 
 When we talk of wave theories, molecular theories, 
laws of attraction, etc., they are nothing independent of 
ourselves but intellectual constructions, necessary to 
simplify the complices of sensation which we call light, 
heat, planetary motion, etc.} 		
Pearson (1880),  Pollock's Spinoza, p.95.
\end{myquote} 

Jacobi's metaphor of Inverted Spinozism presents Transcendental 
Idealism as -- the flaring up of a pure flame, burning all by itself, 
with no need of place or material to nourish it.  
An external cosmos would provide such a place for the 
Transcendental Ego's flame to exist, and  natural laws 
governing material entities in this cosmos would provide this 
flame with a base or nourishment.   
  
 Nowadays, many didactic texts in statistics present K.Pearson 
philosophical positions as a form of mild subjectivism, a subjectivism that, 
on one hand, opposes naive or dogmatic forms of realism but, 
on the other hand,  still allows the scientist to  (in some form)   
relate to the external reality or to connect to existing objects 
in the world-out-there.      
 In other words, a subjectivism that allows the scientist to establish 
some form of harmony between the laws governing the way  objects   
``are'' in the field of study of his science, a  realm that stands by itself, 
and the rules of thought used to think about the same objects.  
 This is absolutely not the position taken by K.Pearson, as it could 
be deduced from either Fichte's definition or Schelling's 
explanation of the expression Inverse Spinozism.

Further explanations concerning the error of seeking some   
harmony between rules of though and laws of nature, 
or between internal representations and external order,     
can be found in Pearson (1897),  Philosophy of Natural Science, 
see following quotations, where K.Pearson criticizes the ideas 
of Paul Volkmann concerning this topic.  
For further details on the physical and epistemological ideas 
of Paul Volkmann, see Corry (2004, p.61-63) and 
Volkmann (1896, 1900).   
  Volkmann is not only a scientist, but also a physics' professor 
greatly concerned with good didactic practices. 
 As such, he comes to the conclusion that teaching physics 
as an axiomatic system is a bad idea, 
for it is neither a faithful depiction of how science is actually built, 
nor a way of presentation that favors its understanding. 
 Instead, Volkmann suggests an iterative approach as the 
best option for both understanding and explaining  the role played 
by the several elements that constitute a scientific theory.  
    
 Volkmann presents his epistemological ideas using the 
metaphor of -- 
\textit{The scientific system as a vaulted arch}: 
 As the stones in a vaulted dome,  the elements of a 
scientific system mutually support each other; 
as the constructive pieces of an arched bridge, 
the elements of a scientific work of architecture are  
kept in place by their mutual interaction.  
 This beautiful metaphor is further explained in Volkmann's 
own words in the following quotation:

\begin{myquote} 
\textit{ 
You can immediately ask for the deeper underlying reason concerning why 
there is  a need, in the physical sciences, for an iterated cycle of knowledge. 
The reason can easily be seen in the form of emergence and 
internal operation of the physical sciences: 
The conceptual system of physics should not be understood as a system  
that is constructed in the manner of a building, from the bottom-up. 
 Rather, it should be understood as a system of cross-references 
that is constructed like a vaulted dome or the arch of a bridge.  
 In this way, it is a system that demands various mutual and 
reciprocal references, in which future results should be known 
from the start and, on the other way around,  several previously 
stipulated conditions must be sustained retrospectively.   
  Physics, in short,  is a conceptual system that is retroactively 
consolidated.}\footnote{\textit{ 
Sie k\"{o}nnen gleich hier nach dem tiefer liegenden Grunde fragen, 
warum f\"{u}r die physikalische Wissenschaft ein wiederholter Kreislauf 
der Erkenntniss nothwendig ist. Dieser Grund ist nach Entstehung 
und Betrieb der physikalischen Wissenschaft einfach darin zu sehen, 
dass das physikalische Begriffssystem nicht etwa aufzufassen ist als 
ein System, welches nach Art eines Geb\"{a}udes 
von unten aufgef\"{u}hrt wird, sondern als ein durch und durch 
gegenseitiges Bezugssystem, welches nach Art eines Gew\"{o}lbes 
oder eines Br\"{u}ckenbogens aufgef\"{u}hrt wird und fordert, 
dass ebenso die mannigfaltigsten Bezugnahmen auf 
k\"{u}nftige Resultate bis zu einem gewissen Grade von 
vorneherein vorweg genommen werden m\"{u}ssen, wie umgekehrt 
bei sp\"{a}teren Ausf\"{u}hrungen die mannigfaltigsten 
Zur\"{u}ckverweisungen auf fr\"{u}here Verf\"{u}gungen und 
Festsetzungen statthaben m\"{u}ssen. Die Physik ist kurz ein 
Begriffssystem mit r\"{u}ckwirkender Verfestigung.}   
 My translation of %
Volkmann (1896, 1910, p.113-114), as restated in Volkmann (1900, p.3-4), 
see also Volkmann (1896, 1910, p.245) and Corry (2004, p.61-63).  }
\end{myquote}

Paul Volkmann's epistemological position is far from any form of 
na\"{\i}ve or dogmatic realism. Instead, perhaps way ahead of his 
time, I would characterize him as a protoconstructivist.  
Nevertheless, Vokmann is also far away of K.Pearson's 
science of the Absolute Ego. 
Accordingly, K.Pearson dissects Volkann's work, 
pin-pointing what he considers to be his fundamental mistake.   
 This fine critical exercise forces him to express his points of view  
in a very clear and candid fashion; this adversarial exercise forces 
K.Pearson to make it abundantly clear how far he is willing to walk 
his path of Inverted-Spinozism or Transcendental Idealism. 
  Specifically, K.Pearson  emphasizes the epistemological error 
of conferring an ontological status or simply acknowledging 
any thing-in-itself  standing alone outside the conceiving Ego, 
an Ego that ultimately conceives everything into existence.     
	According to K.Pearson, any attempt to represent,  to connect to,  
or even to learn from an external order of things in themselves,  
leads to a dead end.

\begin{myquote} 
\textit{ 
We might pass to Dr. Volkmann and show the vagueness of 
his definitions, the unphilosophical character of his epistemology, 
and indicate the danger which arises when loose analogies drawn 
from natural science are applied to other fields of thought. ...   
For him {\em [Volkmann]} natural laws like the law of 
gravitation lie outside us while the conclusions of mathematics 
are thought-laws which lie inside us:   
``Diese Naturnothwendigkeiten ausser uns nirgends in Widerspruch 
treten mit den Denknothwendigkeiten in uns.''}  
[These necessities of nature outside us, nowhere contradict the 
  necessities of thought in us.]\footnote{ 
 Pearson (1896) gives only the German text; 
my translations are added in square brackets.} 	

\textit{ 
This arises apparently from a pre-established harmony the source 
of which is accounted for in a manner which the writer tells us is the  
``Kernpunkt meiner erkenntnistheoretischen Studien auf 
naturwissenschaftlichem Boden.''}  
[The core of my epistemological studies concerning the 
 foundations of natural sciences.]  
  
\textit{It {\em [the harmony]} lies namely in this:   
``dass die Logik in uns ihren Ursprung in dem gesetzm\"{a}ssigen 
Geschehen der Dinge ausser uns hat, dass die \"{a}ussere 
Nothwendigkeit des Naturgeschehens unsere erste und recht 
eigentliche Lehrmeisterin ist.''} 
[That the logic in us has its origin in the regular happenings 
of things outside us, that the external necessity of the 
natural events actually is our genuine and true instructor.]
\textit{ 
We are only given this sentence, without one word more description 
of the process by which such harmony has been established!}   
\end{myquote}

In conclusion, taking Jacobi's metaphor of \textit{Inverted Spinozism} 
in its full extent, we can understand how K.Pearson reached the 
conclusion, see \textit{The Grammar of Science} (1911, p.109-110),  
that the science of Transcendental Idealism can neither be 
an internal version or representation of external laws, 
nor the product of a harmony between internal rules of reasoning 
and external laws governing the cosmos. 
 Instead, so-called laws of nature must be the product of internal 
harmonizations of distinct aspects or faculties within the conceiving Ego.  
 That is, according to K.Pearson, a scientist  should recognize that the 
laws of science result from...

\begin{myquote} 
\textit{
...the harmony between his perceptive and reasoning faculties... 
Thus both the material and the laws of science are inherent 
in ourselves rather than in an outside world. 
 Our groups of perceptions form for us reality, and the results of 
our reasoning on these perceptions and the conceptions deduced 
from them form our only genuine knowledge.}  
\end{myquote}

\subsection{Phenomenological Science vs. Causal Theories}

If it is an error to seek for ``laws of nature'', even if such laws are 
derived from a pre-existent or constructed harmony between 
rules of thought and rules governing objects of an external order, 
then it must be an even greater mistake to ask for causal 
explanations, that is, to seek valid answers to questions asking 
why  ``things''  behave the way they do 
(according to a  given law of nature).   
  The next quotation, from 
	Pearson (1901, The Prostitution of Science, p.50), 
further analyzes the  absurdity  of any metaphysical quest, 
that is, the nonsense of searching for such causal explanations.

\begin{myquote} 
\textit{ To argue from the harmony existing among my sensations 
to a like harmony and order in the {\em Dinge an sich} 
[things-in-themselves] is to multiply 
needlessly the causes of natural phenomena... 
If the human perceptive faculty is capable of so co-ordinating 
sensations that all the groups maintain their own sequence, 
and are in perfect harmony with each other, shortly that `order' 
and `design' appear in natural phenomena,  what advantage do we
gain by needlessly multiplying causes and throwing back the
`order' and harmony of our sensations upon the {\em Dinge an sich},
and an unknowable intellectuial faculty behind them? } 
\end{myquote}

In a very  condescending manner, K.Pearson speaks to 
the addicted to metaphysics, the seekers of causal explanations, 
trying to save them from wandering their erroneous path:  
  
\begin{myquote} 
\textit{ In all these cases we are dealing with the sequences of
various types of motion, into which we analyze or reduce
a variety of sense-impressions. Just as in the special case
of gravitation, we can also describe these sequences and
can frequently give a measure to the motions which we
conceive to take place, but we are still wholly unable to
state why these motions occur. We may talk, if we
please, about the forces 
 [and causes, and parameters]  
... but in using such phrases we do not
introduce an iota of new knowledge, but too often a whole
alphabet of obscurity. We hide the fact that all knowledge
is concise description, all cause is routine.} 
Pearson (1911, Ch.IV Cause and Effect - Probability, p.133). 
\end{myquote}

 Of course, many sensible scientists and philosophers were 
not  so easily convinced to abandon the mission of explaining 
and understanding laws of nature, 
for they knew full well how important causal reasoning can be 
in the process of applying these laws to real problems in their 
fields of expertise, in conceiving new scientific hypotheses or theories, 
in implementing new experiments or research projects, etc. 
 Lord Arthur James Balfour (1894, 1902) was among those totally 
unconvinced by K.Pearson's anti-metaphysical  arguments. 
 K.Pearson (1897) answers to lord Balfour  are very illuminating.  
 Once more, the adversarial style used in this disputation makes 
it  clear how committed he is in following his path of 
Inverted-Spinozism.

\begin{myquote} 
\textit{ Mr. Balfour speaks contemptuously of 
those who regard the Universe as a ``mere collection of 
{hypostatised sense-perceptions} packed side by side in 
space and following each other with blind uniformity 
in time.'' He wants { ``ideas of wider sweep and richer 
content''}, and considers that the work of Science would 
be beneath contempt if it only provided a machinery by 
which the re-occurrence of feelings and ideas might be 
adequately accounted for.} 
Pearson (ChD, p.195). 

\textit{ It does not matter whether it be Spinoza  with his 
Infinite Substance, or Kant with his Dinge-an-sich, 
or the naturalist with his molecule, or the theologian 
with his personal God -- one and all can tell us nothing 
of the real mode of action of his {\em idolum specus}. } 
Pearson (ChD, p.193).

\textit{ The physicist, { who projects his concepts   
into the unknowable} beyond sense-impression, 
is as unphilosophical and as dogmatic as 
the metaphysician or theologian.} 
Pearson (ChD, p.202). 
\end{myquote}

 If the product of scientific research shall never be projected 
outside the conceiving Ego, if science can not go beyond  
accommodations between distinct properties of the perceiving,  
reasoning and conceiving Ego, 
what shall be the purpose of science?  
 According to K.Pearson, the purpose of science is purely descriptive. 
This descriptive function encompasses past and future 
observations, that is, this role of description includes prediction, 
and this must be the sole motivation of scientific research.  
 This point is made unequivocally clear in K.Pearson essay 
answering Balfour objections:  

\begin{myquote} 

\textit{ ...how is Science related to the phenomenal world? 
 Simply as providing comprehensive descriptive formula 
 -- so-called laws -- summed up in a conceptual model 
which more or less completely figures past and 
rehearses future experience. 
 The { symbols of Science} are not ``things in 
themselves,'' nor are they perceptions 
-- nay, as a rule, they do not even stand as 
equivalents for concrete and actual phenomena.}    
Pearson (ChD, p.203).

\textit{ [T]he mission of { Science is not to explain 
but to describe}; to discover a { descriptive formula 
which will enable men to predict} the nature of 
future perceptions; such descriptive formulae are, 
in the only consistent sense of the word, 
knowledge, they form that { ``economy of thought''},   
which is the name happily devised by a philosophical 
physicist to describe and define Science.} 
Pearson (ChD, p.200). 

\end{myquote} 

 This last conclusion can be taken as the net result of 
K.Pearson's approach to philosophy of science. 
 The role played by descriptive/ predictive formulas in  
science offers, on one hand, a touchstone for validation 
criteria and, on the other hand,  avoids 
any metaphysical involvements and complications. 
 This is K.Pearson's key to open the realm of epistemology.  
 K.Pearson's motivations in the philosophy of Spinoza and 
Fichte are today almost forgotten. 
 Nevertheless, the descriptive/ predictive approach to 
scientific hypotheses gained wide acceptance, being developed 
over the first four decades of the XX century into the philosophical 
basis for the Frequentist school of statistics. 
 In latter years, K.Pearson himself gave little emphasis 
to his early undertakings in Transcendental Idealism.   
Instead, he concentrated his efforts to advance the more pragmatical 
thesis of description/ prediction as the ultimate goal of scientific 
hypotheses, and to construct (Frequentist) Statistics as a language 
designed as a tool commissioned to best achieve this goal.    
 The last quotation of this section, stated and restated  by K.Pearson 
in his last years, is typical of this pragmatic approach:

\begin{myquote} 
 \textit{ The Laws of Nature are only constructs of our minds; 
none of them can be asserted to be true or to be false, 
they are good in so far as they give good fits to our observations 
of Nature, and are liable to be replaced by a better `fit'...} 
Pearson (1935) as quoted in Inman (1994, p.6). 
\end{myquote} 				

 For K.Pearson, Goodness-of-fit, and nothing else, becomes the 
ultimate criterion of good science, a criterion that motivates and 
directs the development of Frequentist statistics.  
 In contrast, according to K.Pearson, causal reasoning, laws of 
nature and their explanation, and similar efforts for metaphysical 
understanding have no legitimate role to play in the development 
of contemporary science,  see Section 4 for further comments.

\subsection{Auguste Comte's Positivism and Logical-Positivism} 

 Auguste Comte (1798-1857) had a multifaceted life and  
personality, and so is his philosophical work and legacy,   
see Lacerda (2009), Manuel (1962) and Scharff (1995).   
 As mentioned in the introduction, nowadays many text-books 
in statistics present K.Pearson's philosophy as akin to Positivism. 
 K.Pearson's own opinion was quite different:  
On one hand, he gave Comte the merit of indicating the vanity of causal 
reasoning but, on the other hand, he contended  Comte was never 
able to completely break away from metaphysical forms of thinking. 
 In K.Pearson's candid evaluation of the historical influence 
and importance of Comte's philosophical ideas: 

\begin{myquote} 
\textit{ ...the writings of Comte have at the very least acted as a 
stimulus -- if only of the irritant kind --} 
Pearson (1911; The Grammar of Science, 
p.570).  
\end{myquote} 

 As in previous sections, I use K.Pearson's adversarial arguments to 
highlight and clarify his full commitment  to his Inverted-Spinozism.  
 For K.Pearson, Comte relapses into metaphysics in (at least) 
two crucial occasions, namely, 
(a) the  \textit{Religion of Humanity}; and 
(b) the \textit{Scala Intellectus} schema for classification of 
scientific disciplines. 

(a) The Religion of Humanity is an enterprise of Comte's late life,  
only marginally successful in Comte's native France and some 
other countries. 
 Today the Religion of Humanity has in Rio de Janeiro and 
Porto Alegre, Brasil, its last working temples,  
see Lins (1967) and Valentin (2010).  
 Comte's religious enterprise got him in trouble with many of his 
own followers, and was disregarded without further consideration 
by later versions of Positivism, including the prestigious 
Logical Positivism movement of the Wiener Kreis. 
 For K.Pearson, the Religion of Humanity was not only incompatible  
with his anti-metaphysical principles, but also did not sit well with 
his moral and political ideals.    
  
\begin{myquote} 
\textit{ -- we act morally, that is, socially. 
 Positivism has recognised in a vague impracticable fashion this, 
the only possible basis of a rational morality; it places the progress 
of mankind in the centre of its creed, and venerates a personified 
Humanity.} 
 Pearson (1901; 
 The Moral Basis of Socialism, p.303).  
\end{myquote}

(b) K.Pearson's objections to Comte's \textit{Scala Intellectus} or 
staircase of the intellect schema for classification of scientific 
disciplines  is far more important for the purposes of this paper. 
 In this schema, reminiscent of the Jacob's Ladder studied in 
Stern (2017), scientific disciplines are ranked according to a 
progressive order of complexity of their fields of interest,   
each being indispensable for the study of the next one in 
ascending order.  Specific disciplines are ranked at a 
seven-step-ladder, starting from mathematics and progressing 
to astronomy, physics, chemistry, biology, sociology, and ethics. 
 According  to  Pearson (1892, 1911; The Grammar of Science, 
 The Classification of Sciences,  p.592):   

\begin{myquote} 
\textit{ From Comte [we learn] that there is in reality an 
interdependence in the sciences, so that a clear understanding 
of one may necessitate a previous study of several others.} 
\end{myquote} 

 For K.Pearson, this ``interdependence in the sciences''  
must be unacceptable, for it implicitly opens the door for  
causality relations, even if in concealed forms or in indirect ways, 
between the objects of study of the distinct disciplines, 
following  the very same links according to which these  
disciplines are hierarchically chained into a great chain of being.   
 Furthermore, only getting rid of Comte's artificial hierarchy, can 
K.Pearson allow and justify  crafting of good-fitting descriptive 
formulae for phenomena of interest of a given discipline to be a task 
that can and should be accomplished without recourse to concepts 
of lower ranking disciplines in the same hierarchy.    
 Therefore, it is easy to understand his conclusion stated in  
 Pearson (1892, 1911; The Grammar of Science, 
 The Classification of Sciences,  p.570):   

\begin{myquote} 
\textit{ It is clear that we have in Comte's staircase of the intellect 
[scala intellectus] a purely fanciful scheme, which, like the rest of 
his System of  Positive Polity, is worthless from the standpoint of 
modern science.} 
\end{myquote} 
 
 Finally, it is worth mentioning that, as it should be expected, 
also in the topic at hand (concerning causal chains of being) 
we find K.Pearson's Inverted-Spinozism standing diametrically 
opposed to the philosophy of Spinoza.  
 This contrast is made clear in    
Pearson (1901, Ethics of Renunciation, p.84): 
    
\begin{myquote} 
\textit{
 In his [Spinoza] system, God, we have seen, is identified with the 
reality of things, not things regarded as phenomena, but as links 
in an infinite chain of intellectual causality. 
 He is the $\lambda \textrm{o} \gamma \textrm{o} \varsigma$ 
which dwells in and is all existence;  `laws of nature' 
are only the sensuous expression of the laws of the divine intellect; 
the story of the world is only the phenomenalising of the successive 
steps in the logic of pure thought.} 
\end{myquote} 

This and previous sections' main goal was to understand key aspects  
of K.Pearson's philosophical studies, starting from Spinoza and his 
medieval precursors, and ending at a Fichtean type of Inverted-Spinozism. 
 Furthermore, some consequences of K.Pearson's particular 
version of Inverted-Spinozism have been highlighted by comparison and 
contrast with views and ideas of other thinkers, like Friedrich Jacobi, 
Friedrich von Schelling, Paul Volkmann and Auguste Comte.

\section{From Philosophy to Statistical inference \\ 
	and the Purging of Causal Agencies}
    		
 This section gives a brief overview of K.Pearson work 
in science, from the time of  his spiritual crisis and its 
subsequent philosophical solution to the end of his life. 
 K.Pearson's work in science and philosophy proper 
lost nowadays most of its prestige and influence. 
 Nevertheless,  this section is intended as a bridge, 
providing historical and logical links for the next 
section, dedicated to K.Pearson subsequent 
work in developing the methods and language of  
\textit{classical} or  \textit{frequentist statistics}. 
 Nowadays, statistical science provides the standard accepted 
tools for hypothesis testing and validation, providing 
in this sense a pragmatic logic for scientific inference.    
 Therefore, it is important to analyze and understand how the 
language of classical statistics continues to propagate K.Pearson's epistemological ideas, even if  some of its underlying 
philosophical principles are now discredited, outdated  or just forgotten.

\subsection{Ether Physics and Anti-Atomism} 

 In a series of papers from  1884 to 1891, K.Pearson develops 
some research topics in the area of Ether Physics. 
 Today this area of research is all but forgotten, being only of 
historical interest, see Schaffner (1972) and Whittaker (1953). 
 However, during the XIX century, Ether or Aether physics was 
an important field of study, and a main battle ground between 
the proponents of atomic or molecular theories and their 
antagonists. 
 One of the main goals of Ether physics was to replace 
physical theories explained by effects caused by 
hypothetical  atomic or molecular entities by purely  
phenomenological models based on the hydrodynamic or elastic 
properties of an imponderable medium called Ether.   
  
 K.Pearson  sees the Atom as an archetypical  agent of causality. 
 For example, at physics intra-murus, atoms are agents of causality 
linking the realm of classical mechanics to that of thermodynamics, 
the causal link itself being explained and understood via the tools 
of statistical physics. 
 Extra-murus, atoms and molecules are agents of causality linking 
the discipline of physics to that of chemistry, justifying a link in 
the great chain of being of Auguste Comte's hierarchical schema 
of \textit{Scala Intellectus}. 
 Therefore, K.Pearson's motivations and commitment to the 
development of Ether Physics are easy to understand.  
 Notice that although modern statistical physics was developed 
in the XX century, atomic models explaining thermodynamic 
effects are as old as Robert Boyle's (1627-1691) 
volume-temperature law of gases,  see Brush (1983), 
Needham (2004), Newman (1996, 2006) and Rosenfeld (1953).  

 Unfortunately for K.Pearson, the whole subject of Ether Physics 
was at that time already loosing prestige, until it came to an 
abrupt end in the \textit{Annus mirabilis} of 1905, the 
year of Albert Einstein's papers in \textit{Annalen der Physik} 
concerning Special Relativity, the photoelectric effect and 
Brownian motion. 
 The autopsy of Ether physics reveals two causes for its sudden death:  
 On one hand, Einstein's papers on Special Relativity 
replace Galileo's transformations by Lorentz transformations as 
the fundamental invariance group of  physical theory, 
see Stern (2011) and references therein. 
 This elegant substitution solves and explains with astonishing 
simplicity a host of  problems that the Ether physics program 
was for many decades unsuccessfully trying to circumvent  using 
always more complex and cumbersome phenomenological models.   
 On the other hand,  Einstein's papers  concerning the photoelectric effect 
and Brownian motion, with the support of subsequent experimental work 
by Jean Perrin (1908, 1911) and Robert Millikan (1914, 1916), 
gave sufficient reasons for the overwhelming 
majority of the scientific community to abandon Ether models in 
favor of atomic or molecular theories, for further details and 
references see Stern (2014, 2017).

\subsection{Heredity: Mendelian Genetics vs. Statistical Biometry}

\begin{myquote} 
\textit{In the early twentieth century, the Grammar was understood 
by many, including by Pearson himself, as a philosophical rationale 
for statistics, though in fact he took up statistics only after 
completing its first edition [1892].  
 Thereafter, right to the end of his life, Pearson would make it his 
mission to reshape science using the tools of statistical mathematics. 
 From 1893 to about 1905 he published a series of papers that 
gave a new direction to the field of statistics. In 1901 he founded, 
in collaboration with Francis Galton and W.F.R. Weldon, the journal 
Biometrika, which was dedicated to this project.} 
Porter (2004, p7,8)  
\end{myquote} 


 The short quotation opening this subsection describes the chronology 
of a transition period of K.Person's career. 
 On one hand, his \textit{The Grammar of Science}  
became a resounding success. 
 Far from the romantic and mystical style of \textit{The New Werther}, 
and avoiding the most peculiar eccentricities of Transcendental Idealism, 
this book was written to present his epistemological ideas to the 
working scientist. 
 On the other hand, the scientific research project he had chosen 
to show-case his philosophy, Ether physics, was a complete failure. 
 Therefore, at this point in time, K.Pearson desperately needed 
a new research project that  could be used to demonstrate the 
usefulness of his ideas, and could also provide a refuge for himself 
as a working scientist.  

 K.Pearson was able to find a safe harbor in the fields 
of  biometry, heredity and evolution biology. 
 In 1891 he starts to collaborate with the zoologist Walter 
Frank Raphael Weldon (1860-1906), who soon introduces him to 
Francis Galton (1822-1911) -- from then on K.Pearson's great 
protector,  benefactor and role model in life.   
 A common interest of these scientists was to study the   
inheritance of measurable characteristics in biological populations. 
 How to best approach this problem was at that time a matter of 
great controversy, concerning the existence (or not)  
and the role played by ``genes'' -- (hypothetical) corpuscula 
carrying elementary units of genetic information and/or  
causing the inheritance of specific characteristics. 

 On one hand, the role played by genes in biology is, from an 
epistemological point of view, quite similar to the role played 
by atoms in physics or molecules in chemistry. 
  On the other hand, the chronologies of acceptance of the 
``molecular hypothesis'' in physics and chemistry, and the 
``genetic hypothesis'' in biology are quite different.   
 Indirect evidence for the inheritance of discrete genes controlling 
specific characteristics of biological organisms was available since 
the work of Gregor Johann Mendel (1822-1884).  
 However, Mendel (1865) work, done in seclusion at a Moravian 
monastery, was completely forgotten until his hypotheses 
 postulating discrete genetic coding units 
	were rediscovered by Hugo de Vries (1901, 1903).  
 Even so, the nature of these (at that time hypothetical) genetic 
units and their coding mechanisms remained unknown, 
eluding successive generations of chemists and biologists until 
J.D. Watson and F.H. Crick  (1953) cracked the puzzle of  DNA's  
double-helix structure,  see also Maddox (2003) and Fuller (2003).

 Galton's initial working hypotheses assume the existence of 
gemmules -- genetic corpuscula  transmitting hereditary traits.  
	In order to support this field of research, Galton develops 
key concepts and modeling techniques at the core of 
modern mathematical statistics, including the definition of 
\textit{correlation coefficients} and \textit{linear regression} 
models.  
 For  historical accounts, see Cowan (1972), 
Fancher (1989), Gorroochurn (2016a,b) and Kevles (1985).  
 Galton developed these techniques in order to study the 
simplest non-deterministic connections between cause and effect, 
namely, connections that are normally distributed, linear and 
unidirectional.   
 For his satisfaction, Galton ascertains that such linear models 
produce a very good fit to his populational data banks. 
 Nevertheless, for his great surprise, Galton finds the same linear 
models to be applicable either forwards or backwards in time, 
contradicting his concepts of temporal unidirectionality for 
cause and effect. 
 Galton also realizes that, both conceptually and mathematically, 
such a system could be conceived as driven entirely by a 
Gaussian stochastic process, with no need to an additional 
force or agency causing the system to drift into a preset direction.   
 Consequently,  Galton felt the need for a paradigm shift -- 
abandoning genetic causal theories in favor of  non-causal and 
phenomenological statistical models for biological heredity.

It is at this juncture that Galton and K.Pearson start their 
work together. 
Each man had gone trough a crisis of his own, 
from which they emerged ready to make the same sacrifice: 
Renouncing the bride. 
 After a conceptual crisis, both men were committed to build up a 
science that is purely descriptive-predictive in nature and strictly 
non-metaphysical, with no place for causal entities or explanations.      

 K.Pearson was a talented mathematician and, when the time 
came,  he was ready to help Galton and make progress in the 
development of mathematical statistics.      
 K.Pearson was also a skilful (and sometimes ruthless) 
administrator of human resources. 
 In time, he would regiment many bright minds to work for the cause 
(no pun intended), always running a tight ship and keeping a steady 
course in pursuit of his and Galton's basic programmatic goals.   
 The next quotation, from 1896, presents K.Pearson's  
pledge of allegiance to join forces with Galton. 
  The main objective of following articles  will 
be to show how the Frequentist school of mathematical statistics 
developed its means and methods so that they were tailor made 
to perfectly suit the aforementioned programmatic goals.

\begin{myquote}  \textit{   
A considerable portion of the present memoir will be devoted to the 
expansion and fuller development of Mr. Galton's ideas, particularly 
their application to the problem of bi-parental inheritance. ... 
The causes in any individual case of inheritance are far too complex 
to admit of exact treatment; and up to the present the classification 
of the circumstances under which greater or less degrees of 
correlation between special groups of parents and offspring 
may be expected has made but little progress. 
This is largely owing to a certain prevalence of almost metaphysical 
speculation as to the causes of heredity, which has usurped the 
place of that careful collection and elaborate experiment by which
alone sufficient data might have been accumulated, with a view to 
ultimately narrowing and specializing the circumstances under which 
correlation was measured. 
We must proceed from inheritance in the mass to inheritance in 
narrower and narrower classes, rather than attempt to build up 
general rules on the observation of individual instances. 
Shortly, we must proceed by the method of statistics, 
rather than by the consideration of typical cases.} 
Person (1896, p.255). MCTE 
\end{myquote}

\subsection{Eugenics: Scientific \& Individual Utility, Racial Control}   		

One of the traditional goals and most exalted accomplishments 
of science is to give mankind understanding about the world, 
to provide deep insights on how it works, or some intuition  
on why it is the way it is.  
Not for K.Pearson:  His conception of science is strictly 
phenomenological, its only goal being accurate description 
and prediction of  sense-impressions. 
But if not for wisdom and understanding, what is science worth?

\begin{myquote}
\textit{I am afraid I am a scientific heretic -- an outcast from 
the true orthodox faith -- I do not believe in science for its 
own sake.  I believe only in science for man's sake. ... 
 The first condition for State support is that we [anthropologists] 
show our science to be utile by turning to the problems of racial 
efficiency, of race-psychology, and to all those tasks that 
Galton described as the first duty of a nation --  `in short, to make 
every individual efficient both through Nature and by Nurture.'} 
K.Pearson  (1920, pp.136, 148) 
\end{myquote} 
  
 Pearson's conception of science is utilitarian to the core. 
 In the case of his scientific programs in biometry and heredity, 
the intended use was eugenics:  
    
\begin{myquote} 
\textit{The term National Eugenics is here defined as the study 
of the agencies under social control that may improve or impair 
the racial qualities of future generations either physically or mentally.} 
K.Pearson (1930, p.222). 
\end{myquote} 

 The eugenics project reveals a secondary motivation for 
K.Pearson, Galton and the biometics school dislike of 
heredity theories based on hidden causal entities,  like genes. 
  In such theories, the manifested characteristics of 
an individual, its phenotype, are derived from its 
genotype, expressed through complex non-linear interactions  
of multiple genes, in a process that may be further coupled with 
epigenetic factors. 
 In the context of a genetic theory of heredity, a responsible 
manipulation of population genetics would require  
deep understanding of highly complex genomic networks --  
also taking into account the ecological effects of genetic 
diversity and many other aspects far beyond XX century 
science,  for further comments, see Kevles (1985, p.145). 
  Galton and K.Pearson's approach dismissed the aforementioned 
obstacles  using the following  hyper-simplified definition of heredity.

\begin{myquote}
\textit{Heredity:  Given any organ in a parent and the same or any 
other organ in its offspring, the mathematical measure of heredity 
is the correlation of these organs for pairs of parent and offspring. 
 The word organ here must be taken to include any characteristic 
which can be quantitatively measured.
If tile organs are not those of parent and offspring, but of any two 
individuals with a given degree of blood relationship, the correlation 
of the two organs will still be the proper measure of the strength 
of heredity for the given degree of relationship.} 
Pearson (1896b p.259).    
\end{myquote} 
  
  K.Pearson's strictly phenomenological and uninhibitedly utilitarian 
	view of science may have been one of the contributing factors that 
	lead him to pledge his support to most unscrupulous political projects,    
	like in the following quotation from a speech from 1934 delivered	
	at University College, London.   
	 Pearson died two years later, in 1936. 
	Had he lived a few years longer, he may have recognized  
	in this political project the hand of Mephistopheles -- 
	the scatterer and forgerer of lies -- the fictional author of 
	The New Werther,  the book of romantic dreams of his youth.

\begin{myquote}
\textit{ The Royal Society Council having passed a resolution that
mathematics and biology should not be mixed, Biometrika was founded
with Galton as consultant and Weldon and myself as joint editors.
Buccaneer expeditions in too many fields followed; fights took place
on many seas, but whether we had right or wrong, whether we lost or
won, we did produce some effect. The climax culminated in Galton's
preaching of Eugenics and his foundation of the Eugenics
Professorship. Did I say `culmination'?  No, that lies rather in the
future, perhaps with {\em Reichskanzler} Hitler and his proposals to
regenerate the German people.  In Germany a vast experiment is in
hand, and some of you may live to see its results.  If it fails it will not be
for want of enthusiasm, but rather because the Germans are only just
starting the study of mathematical statistics in the modern sense!}  
K.Pearson (1934, p.23), also quoted in Senn (2003, p.144, 238). 
\end{myquote}

\section{Inverted-Spinozism Lives! \\  
  The Language of  Classical Statistics}

 The main objective of this section is to indicate 
how K.Pearson's  philosophy ultimately engenders 
the epistemological foundations of  
\textit{classical} or \textit{frequentist  statistics} 
(XX century mainstream school), strongly influences the 
evolution of the theoretical framework of statistical science in 
general (directly for the frequentist school, and indirectly for 
the Bayesian school), and imposes implicit guide-lines  for  the 
development of well-conforming operational methods and models. 
	Moreover, the language and dialects developed 
by these statistical schools were specifically designed to facilitate 
the acceptance of K.Pearson's philosophical principles in the practice 
of science, and to induce ``epistemologically correct''  forms of 
expression and communication in scientific research. 
 
 Although many of K.Pearson's philosophical and scientific ideas 
are now considered obsolete or discredited,  the language of 
classical statistics, crafted by him and his co-workers,  
not only survives but thrives up to the present days.  
  Since the second half of the XX century, statistical significance 
measures (following the nomenclature of frequentist statistics) 
became the accepted standard by which scientific 
hypotheses must be judged. 
 Therefore, we can understand how K.Pearson's   
philosophy ultimately spreads its influence far and wide, 
even if rigorous philosophical principles got diluted 
along the way, hence exerting their influence in subtle, 
sometimes almost subliminal forms. 
 Indeed, on one hand, mainstream contemporary statistics clearly 
upholds the descriptive/ predictive and non-explanatory nature of  
statistical models in particular and of science in general 
while, on the other hand,  downplays  the historical origins and 
epistemological foundations of the same philosophy, rooted in 
K.Pearson's version of Inverted-Spinozism.    
  In the language of statistics, K.Pearson's Inverted-Spinozism 
is translated as deprecation of inverse-probabilities, 
as explained in the following paragraphs.

 Statistical models distinguish two classes of variables, 
namely: 
On one hand, there are variables in the sample-space.  
These variables are associated with observable 
phenomena, quantities of interest to which we assign direct probabilities. 
On the other hand, there are variables in the parameter-space.  
These variables are associated with latent or non-observable quantities that often correspond to hidden causes of the observed phenomena. 
The most basic learning mechanism in statistical models,   
\textit{Bayes rule}, is a formula used to update inverse  
probabilities according to newly available observations.

  Bayes rule was first described in the article    
\textit{An Essay towards solving a problem in the Doctrine of 
Chances}, published at 
 The Philosophical Transactions of the Royal Society in 1763. 
 In this article, the ideas of the Presbyterian minister Rev. 
Thomas Bayes (1701-1761) were communicated posthumously 
by Rev. Richard Price (1723-1791), nominated by Bayes 
as his literary executor. 
The motivations for Bayes work are clearly stated  
in his work, see next quotation.  
At the time, the expression \textit{inverse-probability} 
was not yet in use, instead, since parameters often  
stand for hidden causes for observed phenomena, 
the expression \textit{probabilities of causes} 
was commonly used, see Fienberg (2006). 
    
\begin{myquote}  \textit{  
 The purpose, is to shew [show] what reason we have for 
believing that there are in the constitution of things, 
{fixed laws} according to which events happen, and that, 
therefore, the frame of the world must be the effect of the 
{wisdom and power of an intelligent cause}; 
and thus to confirm the argument taken from final causes for 
the existence of the Deity [...and...]  
it will be easy to see that the problem solved in this essay is 
more directly applicable to this purpose; for it shews [shows] 
us, with distinctness and precision, in every case of any particular order or recurrency of events, what reason there 
is to think that such recurrency or order is derived from 
{stable causes or regulations in nature}, and not from 
any of the irregularities of chance.} 
Bayes (1763), as quoted in Barker (2001, p.84).  
\end{myquote}

 Reading Bayes explanations, as presented by Price, 
is seems that their research perfectly conforms to  
the Spinozian program of 
\textit{cognitione causae et leges naturae universales}.  
In subsequent generations,  probabilists like 
 Pierre-Simon de Laplace (1749-1827) and 
George Boole (1815-1864) champion the efforts for 
modeling causal relations and computing probabilities 
of possible causes, developing the mathematical 
techniques needed to build and formally present 
the first workable statistical models.  
 As explained in Sections 2, 3 and 4, this course of 
development was turned around by the work of K.Pearson.

Methodologically, the frequentist school can be characterized by 
allowing the use of direct probability statements, that is, 
by considering observables as random variables, 
while strictly forbidding inverse probability statements, 
that is, by never considering random variables 
in the parameter-space. 
The Frequentist school's deprecation of inverse-probabilities is 
a 180$^\circ$ turn, a complete reversal of a long-standing tradition 
in the history of probability and statistics, for inverse probability 
methods had been developed by leading figures of preceding 
generations, like Bayes, Laplace and Boole.  
     
 The most powerful weapon K.Pearson developed for 
fighting his anti-metaphysical war was a new language, 
the language of the \textit{frequentist} school 
(nowadays also knows as the \textit{classic} school)  
of  mathematical statistics. 
 In this development effort, K.Pearson had to fight in 
internal and external fronts.  
 Externally, he had to promote ``classical''  statistics as the 
language of choice for the analysis and validation of scientific 
research. 
 Internally, he had to battle heresies inside the statistical 
community, and assure the crafting of a language that is 
conducive of orthodox thinking, expressing arguments that 
are, by construction, free of metaphysical contamination. 
 These battles are the primary focus of forthcoming articles, 
as further explained in the next section.

\section{Final Comments and Further Research} 

 In this section present some final comments,    
propose some additional topics for further research,    
and give due acknowledgments.

Spinoza, Bayes, Price,  Comte, Boole and K.Pearson, 
all allowed their religious believes to inspire their 
philosophical or epistemological principles that, 
in turn, directly motivated their scientific work and ideas. 
 The study of historical interfaces and influences between 
science and religion seems to have gone out of fashion, 
with scientists and academic philosophers and theologians 
often in accordance about the necessity to separate their 
domains of study,  an isolation effort intended to protect 
each specialty from possible harmful interference.   
 The topics discussed in the present article and in Stern (2017) make me seriously doubt the viability and validity of 
artificially imposing such hermetic seals.   
 Further research in the following topics should help 
to clarity this position.

\subsection{Statistical Languages and Philosophical Commitments}   

 Following articles will show how K.Pearson an his 
coworkers cast the language of mathematical statistics 
in order to enforce  the tacit acceptance and practical 
use of his philosophical and epistemological ideas. 
 Several key choices in the construction of modern 
statistics can not be explained solely by logical 
properties and mathematical requirements. 
 Instead, these key choices can only be understood as reflecting 
epistemological principles based on K.Pearson's inverted-spinozism, 
even if at the expense of developing a formalism with fewer or 
weaker  functionalities and inconsistent logical properties, 
see Borges and Stern (2007), Esteves et al. (2016), 
Madruga (2001), Pereira et al. (2008), Stern (2017, 2018), 
Stern and Pereira (2014), and Stern et al. (2017).    
 Moreover, following articles will show how K.Pearson managed  
to be very successful in his campaign, using the language of 
frequentist statistics to promote and consolidate the 
acceptance not only of his  statistical methodologies, 
but also of his underlying (but far less advertized) 
philosophical principles.

\subsection{Frequentist Epistemology and Logical Positivism}  

 Later versions of the Positivist movement, known as 
Logical Positivism, Empirical Positivism or Logical Empiricism, 
were developed under the influence of the Wiener Kreis 
from 1907 to 1938, and subsequently around the world. 
 These neopositivist programs abandon the goal of becoming 
all encompassing systems of philosophy, taken instead a more 
pragmatic approach, concentrating their efforts into 
specific epistemological questions and the development of 
mathematical logic as a tool to support their research program.  

  Rudolf Carnap (1932)  stated as a main goal of the neopositivist 
program to be -- \textit{The Elimination of Metaphysics Through 
Logical Analysis of Language}. 
 Meanwhile, K.Pearson and his school were developing Frequentist 
statistics as a tool for building up a science free of metaphysics.  
 This is a far more ambitious and radical proposal since statistics, 
as conceived by K.Pearson, is not a tool designed to analyze 
science as it is and circumvent its metaphysical problems, 
but a tool designed for actually making scientific research in  such 
a way that science becomes, by construction, free of metaphysics. 

 I have the impression that the two approaches, based on their 
respective supporting languages, namely, mathematical logic 
and mathematical statistics, had the potential of greatly 
benefiting from each other. 
 However, the interaction between researchers of the 
two groups and their mutual influence seems to have been 
rather small.  
 Nevertheless, in the second part of the XX century, science 
practitioners and writers of didactic textbooks have often 
confused ideas coming from neopositivism and Frequentist 
statistics, carelessly overlapping topics that would require 
more careful treatment.    
 I believe that confirming these first impressions and trying to 
understand these parallel courses of historical development 
are topics that deserve future research.

\subsection{Acknowledgements}   

The author is grateful for the support of IME-USP --   
the  Institute of Mathematics and Statistics 
of the University of S\~{a}o Paulo;  
FAPESP -- the State of S\~{a}o Paulo Research Foundation  
(grants CEPID 2013/07375-0 and CEPID 2014/50279-4);  
and CNPq -- the Brazilian National Counsel of Technological and 
Scientific Development (grants PQ 301206/2011-2 and PQ 301892/2015-6).  
Finally, the author is grateful for advice and comments concerning 
the first version of this paper received from Rafael Bassi Stern, 
Rabbi Ruben Sternschein, Jean-Yves B\'{e}ziau, from anonymous referees,   
from participants of the  2nd World Congress on Logic and Religion, held on June 18-22, 2017, at Warsaw, Poland, 
and from participants of the X Principia International Symposium, held on August 14-17, 2017, at Florian\'{o}polis, Brazil.

 \section*{References} 
 
 \renewcommand{\baselinestretch}{0.89}
 \parskip 0.70mm 
 \begin{small} 



\rr F. Barker, R. Evans (2001). 
\textit{The Probability of Mr Bayes: A constructive re-evaluation 
of Mr Bayes' essay and of the opinions concerning it expressed
by various authorities}. Technical report, Department of
Electrical and Electronic Engineering, University of Melbourne.  

\rr T. Bayes (1764). An Essay Towards Solving a 
Problem in the Doctrine of Chances. 
Phil. Trans. Roy. Soc. London, 53, 370-418. 

\rr W.Borges, J.M. Stern (2007).  The Rules of Logic 
Composition for the Bayesian Epistemic e-Values. 
{\it Logic Journal of the IGPL}, 15, 5-6, 401-420. 

\rr S.G. Brush (191883). \textit{Statistical Physics and the Atomic 
Theory of Matter, from Boyle and Newton to Landau and Onsager}. 
Princeton Univ. Press.  

 \rr R. Carnap (1932). \"{U}berwindung der Metaphysik durch 
 Logische Analyse der Sprache. \textit{Erkenntnis}, 2, 1932. 

\rr M. Clark (1999). \textit{Etymological Dictionary of Biblical 
Hebrew: Based on the Commentaries of Samson Raphael Hirsch}. 
Jerusalem: Feldheim Publishers. 

\rr L. Corry (2004). \textit{  David Hilbert and the Axiomatization of Physics 
(1898--1918): From Grundlagen der Geometrie to Grundlagen der Physik}. 
Dordrecht: Springer. see p.61--63 on Paul Volkmann.  






\rr R.S. Cowan (1972). Francis Galton's Statistical Ideas: 
The influence of eugenics. \textit{Isis} 63,4, 509-528.

   

\rr O. Darrigol (2000). \textit{Electrodynamics from Ampere to Einstein}. 
NY: Oxford University Press. 

\rr A. Einstein (1905a). \"{U}ber einen die Erzeugung und 
Verwandlung des Lichtes betreffenden heuristischen Gesichtspunkt 
[On a Heuristic Point of View about the Creation and Conversion of Light].   
\textit{Annalen der Physik}, 17, 6, 132-148. 

\rr A. Einstein, (1905b). \"{U}ber die von der molekularkinetischen Theorie 
der W\"{a}rme geforderte Bewegung von in ruhenden Fl\"{u}ssigkeiten 
suspendierten Teilchen [Investigations on the theory of Brownian Movement]. 
\textit{Annalen der Physik}, 17, 8, 549-560. 

\rr A. Einstein (1905c). Zur Elektrodynamik bewegter K\"{o}rper 
[On the Electrodynamics of Moving Bodies]. 
\textit{Annalen der Physik}, 17, 10, 891-921. 

\rr A. Einstein (1905d). Ist die Tr\"{a}gheit eines K\"{o}rpers von seinem 
Energieinhalt abh\"{a}ngig? [Does the Inertia of a Body Depend Upon 
Its Energy Content?].  \textit{Annalen der Physik}, 18, 13, 639-641. 

\rr L.G.Esteves, R.Izbicki, R.B.Stern, J.M.Stern (2016). 
The Logical Consistency of Simultaneous Agnostic Hypothesis 
Tests. \textit{Entropy}, 18, 256-278. 

\rr R.E. Fancher (1989). Galton on examinations: An unpublished 
 step in the invention of correlation. \textit{Isis}, 80, 3,  446-455. 

\rr S.E. Fienberg (2006). When Did Bayesian Inference Become
``Bayesian''? \textit{ Bayesian Analysis}, 1, 1, 1-40. 

\rr C. Fraenkel (2006). Maimonides' God and Spinoza's Deus sive Natura
 \textit{ Journal of the History of Philosophy}, 44, 2, 169-215.   

\rr M.L.von Franz (1970, 1990). \textit{Individuation in Fairy Tales}. 
 Boston: Shambhala Publications.  

\rr M.L.von Franz (2000). \textit{The Problem of the Puer Aeternus}. 
 Toronto: Inner City Books. 

\rr Z. Frayzyngier (1979). Notes on the R1R2R2 Stems in Semitic. 
\textit{Journal of Semitic Studies}, 24, 1,  1-12.    

\rr W. Fuller (2003).  Who said `helix'?  
 \textit{Nature} 424, 6951,  876-878.

\rr F. Galton (1869). \textit{Hereditary Genius}. London: Mcmillan. 

\rr  F. Galton (1877). Typical laws of heredity. 
\textit{Nature}, 15, 492-495, 512-514, 532-533. 


\rr F. Galton, (1885a). Presidential Address, Section H, Anthropology. 
\textit{Nature} 32, 507-510. 

\rr F. Galton (1885b). Regression towards mediocrity in hereditary 
stature.  \textit{Journal of the Anthropological Institute}, 15, 246-263.

\rr F. Galton (1988). Co-relations and Their Measurement. 
\textit{Proceedings of the Royal Society},4, 5, 135-145. 

\rr F. Galton (1889).  \textit{Natural Inheritance}. 
      London: Macmillan. 

\rr F. Galton (1904). Eugenics: Its Definition, Scope, and Aims. 
The American Journal of Sociology, X, 1. 

\rr W. Gesenius, E. Kautzsch, A.E.Cowley (1910). 
\textit{Gesenius' Hewrew Grammar}. Oxford: Clarendon Press. 

\rr W. Gesenius (1906). \textit{A Hebrew and English Lexicon 
of the Old Testament}. Boston: Houghton, Mifflin and Co. 

\rr  P. Gorroochurn (2016a). On Galton's Change From ``Reversion'' 
 to ``Regression''. \textit{The American Statistician}, 70, 3, 227-231.
   
\rr  P. Gorroochurn (2016b).  \textit{Classic Topics on the History of 
Modern Mathematical Statistics: From Laplace to More Recent Times}. 
NY: Wiley. 	
	
	
\rr W.Z. Harvey (1981). A Portrait of Spinoza as a Maimonidean. 
\textit{ Journal of the History of Philosophy}, 19, 2, 151-172.  

\rr W.Z. Harvey (2007). Idel on Spinoza. \textit{ Journal for 
 the Study of Religions and Ideologies}, 6, 18, 88-94.  

\rr W.Z. Harvey (2012). 
 Gersonides and Spinoza on Conatus. 
 \textit{ Aleph: Historical Studies in Science and Judaism},  
12, 2, 273-297. 

\rr W.Z. Harvey (2014).  `Ishq hesheq and 
Amor Dei Intellectualis. p.96-107 in S. Nadler (2014). 
\textit{ Spinoza and Medieval Jewish Philosophy}, 
Cambridge Univ. Press. 

\rr E. Horowitz (1993). 
\textit{How the Hebrew Language Grew}. Jerusalem: KTAV. 

\rr A.J. Howell (2015). Finding Christ in the Old Testament 
through the Aramaic Memra, Shekinah and Yeqara of the 
Targums.  Ph.D. Thesis, Louisville, KY: 
The Southern Baptist Theological Seminary. 

\rr M. Idel (2000). Deus Sive Natura -- The Metamorphosis of a 
Dictum from Maimonides to Spinoza. pp.87-110 in 
R.S. Cohen, H. Levine (2000). \textit{ Maimonides and the Sciences}. 
Boston Studies in the Philosophy of Science, v.211. 

\rr M. Idel (2000). \textit{ Maimonide e la Mistica Ebraica}. 
     Genova: Il Melagolo.  

\rr M. Idel (1988). \textit{ Studies in Ecstatic Kabbalah}. 
    Albany: SUNY Press. 

\rr M. Idel (2005). \textit{ Kabbalah and Eros}. New Haven.  
	
\rr  Jacobi (1799).  Jacobi an Fichte (open letter). Hamburg: Perthes. 
Translated in p.497-536 of  F.H. Jacobi (1994). 
Main Philosophical Writings and the Novel Allwill. 
Mcgill Queens Univ Press. 


\rr D.J. Kevles (1985). \textit{In the Name of Eugenics: Genetics and 
the Uses of Human Hereditarity}.  Univ. of California Press. 

\rr E. Klein (1987). \textit{A Comprehensive Etymological Dictionary 
of the Hebrew Language}.  Jerusalem: CARTA. 
   

	\rr G. Levine (2010). \textit{Dying to Know: Scientific Epistemology 
	and Narrative in Victorian England}. Univ. of Chicago Press. 


\rr G.B. de Lacerda (2009). 
Augusto Comte e o Positivismo Redescobertos. 
 \textit{Revista de Sociologia e Pol\'{\i}tica}, 17, 34, 319-343.

\rr I. Lins (1967). \textit{Hist\'{o}ria do Positivismo no Brasil}. 
Brasiliana, v.322, S\~{a}o Paulo: Companhia Editora Nacional.  	
   
\rr B. Maddox (2003). The double helix and the `wronged heroine'. 
 \textit{Nature}, 421, 6921,  407-408.	

  \rr  M.R.Madruga, L.G.Esteves, S.Wechsler (2001).
 On the Bayesianity of Pereira-Stern Tests.
 {\it Test}, 10, 2, 291-299.
	
\rr  M. Maimonides, transl. H.H. Bernard (1832).  
\textit{Yad HaChazakah: The Main Principles of The Creed And Ethics 
of The Jews}. Cambridge: J.Smith.  

\rr F.E. Manuel (1962). \textit{ The Prophets of Paris: Turgot, Condorcet, 
Saint-Simon, and Comte}. NY: Harper and Row. 

\rr R. de Marrais (1974). The Double-Edged Effect of Sir Francis Galton: 
A Search for the Motives in the Biometrician-Mendelian Debate. 
\textit{Journal of the History of Biology}, 7, 1, 141-174.

\rr G. Mendel (1865). Versuche \"{u}ber Pflanzen-Hybriden.  
  \textit{Verhandlungen des naturforschenden Vereines in Br\"{u}nn},  
	 IV, 3-47.  

\rr P. Merlan (1963).   \textit{Monopsichism, Mysticism, 
Metaconsciousness:  Problems of the soul in the neoaristotelian and 
neoplatonic tradition}.  The Hague: Martinus Nijhoff. 

\rr R. Millikan (1914). A Direct Determination of h.  
 \textit{Physical Review}, 4, 1, 73-75. 

\rr R. Millikan (1916). A Direct Photoelectric Determination of Planck's h. 
 \textit{Physical Review}, 7, 3, 355-388. 

\rr S.A. Mulaik (1985). Exploratory Statistics and Empiricism. 
\textit{Philosophy of Science}, 52, 3, 410-430. 


\rr P. Needham (2004). When did atoms begin to do any 
explanatory work in chemistry? \textit{International Studies 
in the Philosophy of Science}, 18, 199-219.

\rr W. Newman (1996). The Alchemical Sources of 
Robert Boyle's Chemistry and Corpuscular Philosophy. 
\textit{Annals of Science}, 53. 567-85.
   
\rr W. Newman (2006). Atoms and Alchemy. 
Chicago: University of Chicago Press. 

   
\rr  K. Pearson (1880). Pollock's Spinoza. 
 Cambridge Review, 2, 94-96. 

\rr K. Pearson (1880). The New Werther, 
 by Locki.  London: C. Kegan Paul \& Co.

\rr K. Pearson (1880). Pollock's Spinoza. 
 \textit{Cambridge Review}, 2, 94-96. 

\rr K. Pearson (1883). Maimonides and Spinoza. 
 Mind 8, 338-353.

\rr K. Pearson (1883). Kuno Fischer's New Critique of Kant. 
    \textit{Cambridge Review}, 5, 109-111.

\rr K.Pearson (1884). On the Motion of Spherical and Ellipsoidal 
Bodies in Fluid Media. \textit{The quarterly journal of pure and 
applied mathematics}, 20, 184-192. 

\rr K. Pearson (1885, 1889). On a Certain Atomic Hypothesis. 
\textit{Transactions of the Cambridge Philosophical Society}, 14, 71-120.

\rr K. Pearson (1888, 1889). On a Certain Atomic Hypothesis. 
\textit{Proceedings of the London Mathematical Society}, 20, 38-63. 
				
\rr K. Pearson (1891). Ether Squirts. 
 \textit{American Journal of Mathematics}, 13, 4, 309-362. 						

\rr K. Pearson (1892, 1911). The Grammar of Science. 
      3rd ed., London: Adam \& Charles Black. 	

\rr K. Pearson (1896). Review of P. Volkmann (1896). 
\textit{Nature}, 55, 1410, 1-4.  

\rr K. Pearson (1896b). Mathematical Contributions to the Theory 
of Evolution III: Regression, Heredity, and Panmixia. 
\textit{Philosophical Transactions of the Royal Society of London}, 
Series A, 187, 253-318. 

\rr K. Pearson (1897a). Replica and treplica  to the review of 
P. Volkmann (1896).  \textit{Nature}, 55, 1424, 342-343.   
			
\rr  K. Pearson (1897b). Reaction: A Criticism of 
Mr. Balfour's Attack on Rationalism. p.173-225 in: 
\textit{The Chances of Death and Other Studies in Evolution}. 
London: Edward Arnold.  
	
\rr K. Pearson (1901). \textit{The Ethic of Freethought}. 
 London: Adam and Charles Black. 

\rr K. Pearson (1901). On lines and planes of closest fit
to a system of points in space. 
\textit{Philosophical Magazine}, 2, 557-572.

\rr K.Pearson (1930). \textit{The Life, Letters and Labours of Francis 
Galton --  vol. 3A,  Correlation, Personal Identification and Eugenics}. 
Cambridge University Press. 

\rr K.Pearson (1920). Address to the Anthropological Section. 
Report of the Eighty-Eight Meeting of the British Association 
for the Advancement of Science, Cardiff.  p.134-151.

\rr K. Pearson (1934). \textit{Speeches delivered at a dinner held 
in University College London in honour of Professor Karl Pearson,  
23 April 1934}.  Cambridge University Press.  

 \rr C.A.B.Pereira, S.Wechsler, J.M.Stern (2008). 
 Can a Significance Test be Genuinely Bayesian? 
 {\it Bayesian Analysis,} 3, 1, 79-100. 
   
\rr  J. B. Perrin (1909). Mouvement Brownien et R\'{e}alit\'{e} 
 Mol\'{e}culaire. \textit{Annales de Chimie et de Physiqe}, 
 VIII, 18, 5-114. 

\rr J.B.Perrin (1913). \textit{Les Atomes}. Paris: Alcan. 
      Translation: Atoms. NY: Van Nostrand.
	
	\rr L. Roth (1924, 1963). \textit{Spinoza, Descartes and Maimonides}. 
     NY: Russell \& Russell.  
			
\rr L. Rosenfeld (1953, 2005). \textit{Classical Statistical Mechanics}. 
CBPF - Centro Brasileiro de Pesquisas F\'{\i}sicas. 

\rr R.C. Scharff (1995).  \textit{Comte after Positivism}. 
     Cambridge Univ. Press. 
			
\rr K.E. Schaffner (1972). \textit{Ninteenth-Century Aether Theories}. 
 Oxford: Pergamon Press.

\rr Schelling (1994). On the History of Modern Philosophy.  
Cambridge University Press.  
Transl. from \textit{Geschichte der neueren Philosophie}, 
 Vorlesungen an der Universit\"{a}t M\"{u}nchen, 1827. 

\rr S. Senn (2003). \textit{Dicing with Death: Chance, Risk and Health}. 
Cambridge University Press.  




\rr B. Spinoza (1677, 2008). \textit{Ethica: Ordine Geometrico 
 Demonstrata  /  \'{E}tica: Demonstrada Segundo a Ordem 
 Geom\'{e}trica};  Edi\c{c}\~{a}o Bil\'{\i}ng\"{u}e Latim - 
 Portug\^{e}s. S\~{a}o Paulo: Aut\^{e}ntica.

\rr J.M. Stern (2011). Symmetry, Invariance and Ontology in 
Physics and Statistics. \textit{Symmetry}, 3,  611-635. 

\rr J.M. Stern (2014). 
Jacob's Ladder and Scientific Ontologies. 
{\it Cybernetics \& Human Knowing}, 21, 3, 9-43. 

\rr J.M. Stern (2017). Continuous versions of Haack's 
Puzzles: Equilibria, Eigen-States and Ontologies. 
{\it Logic Journal of the IGPL}, 25, 4,  604-631. 

\rr J.M. Stern (2017). Jacob's Ladder: Logics of Magic, 
Metaphor and Metaphysics. \textit{Sophia} -- on line 
{\tt DOI 10.1007/s11841-017-0592-y}    

\rr  J.M. Stern (2018). Verstehen (causal/ interpretative 
understanding), Erkl\"{a}ren (law-governed description/ 
prediction), and Empirical Legal Studies. 
{\it Journal of Institutional and Theoretical Economics  
-- Zeitschrift f\"{u}r die Gesamte Staatswissenschaft}, 
174, 1, 105-114.  

\rr J.M. Stern, R. Izbicki, L.G. Esteves, R.B. Stern (2017).  
Logically-Consistent Hypothesis Testing and the Hexagon 
of Oppositions. {\it Logic Journal of the IGPL}, to appear. 

\rr J.M. Stern and C.A.B. Pereira (2014). Bayesian Epistemic 
Values: Focus on Surprise, Measure Probability! 
{\it Logic Journal of the IGPL}, 22, 2, 236-254.




\rr O.F. Valentin (2010). \textit{O Brasil e o Positivismo}. 
     Rio de Janeiro: Publit.   

 
        
\rr P. Volkmann (1896).  \textit{Erkenntnistheoretische Grundz\"{u}ge 
der Naturwissenschaften und ihre Beziehungen zum Geistesleben der 
Gegenwart}.  Leipzig:  Teubner, 2nd ed. Berlin 1910.  

\rr  P. Volkmann (1900).  \textit{Einf\"{u}hrung in das Studium der 
theoretischen Physik, insbesondere das der analytischen Mechanik, 
mit einer Einleitung in die Theorie der physikalischen Erkenntinis}. 
Leipzig: Teubner, 2nd ed. Berlin 1913. 

\rr H. de Vries (1901, 1903). \textit{Die mutationstheorie.  Versuche und 
  beobachtungen \"{u}ber die entstehung von arten im pflanzenreich}. 
	 Leipzig: Veit. 

\rr J.D. Watson,  F.H. Crick  (1953). Molecular Structure of Nucleic 
 Acids: A Structure for Deoxyribose Nucleic Acid. 
 \textit{Nature}, 171, 4356, 737-738.  
				
\rr E.Whittaker (1953). \textit{History of the Theories of Aether 
and Electricity}. London:  Longmans. 

 \end{small} 						
 


\

\authorname{Julio Michael Stern}
\address{%
	Institute of Mathematics and Statistics of the University of S\~{a}o Paulo. \\ 
	Rua do Mat\~{a}o 1010,  CEP 05508-090, S\~{a}o Paulo, Brazil. }
\email{jstern@ime.usp.br, jmstern@hotmail.com}

\end{document}